\documentclass{article}
\usepackage {graphicx}
\begin{document}
\baselineskip .75cm 
\begin{titlepage}
\title{\bf Virial expansion and condensation with a new generating function}       
\author{Vishnu M. Bannur  \\
{\it Department of Physics}, \\  
{\it University of Calicut, Kerala-673 635, India.} }   
\maketitle
\begin{abstract}

Mayer's convergence method for virial expansion and condensation is studied using a new generating function for canonical partition function, which directly depends on irreducible cluster integral, $\beta_k$, unlike Mayer's work where it depends on reducible cluster integral, $b_l$. The virial expansion,  criteria for it's validity and criteria for condensation, etc. are derived from our generating function. All earlier Mayer's results are obtained from this new generating function. 
\end{abstract}
\vspace{1cm}
                                                                                
\noindent
{\bf PACS Nos :}  05.20.-y, 05.70.Fh, 64.60.-i, 64.70.F-\\
{\bf Keywords :} Classical cluster expansion, virial expansion, condensation.
\end{titlepage}
\section{Introduction :}

Many real systems are classified as non-ideal such as gas and liquid \cite{ma.1}, high density plasma \cite{bl.1, za.1}, quark gluon plasma \cite{ba.1,bi.1,ud.1,el.1}, etc. When the interaction between the constituents become strong enough, non-ideal effects become important and leads to the problem of phase-transition. The problem of phase-transition is still unsolved problem in statistical mechanics. Only the approximate system, mimicking the observed phase-transition, may be studied using models like Heisenberg, Ising, etc \cite{pa.1}  and explains qualitative features near the critical point.  

In this article, we study the Mayer's theory of virial expansion and condensation in the light of recent work of Ushcats \cite{us.1}. Following the definitions and notations of Pathria \cite{pa.1} on classical cluster expansion, here we derive a new generating function for canonical ensemble partition function. Using this generating function which directly depends on irreducible cluster integrals $\beta_k$, we study the  Mayer's theory. In Mayer's procedure \cite{ma.1} one starts from a generating function which depends on the cluster integral $b_l$ and one needs one more relation connecting cluster integrals $b_l$ and $\beta_k$ to discuss virial expansion and condensation.   

First we review the Mayer's theory of virial expansion in section 2. In section 3, yet another derivation of virial expansion is discussed. Mayer's convergence theory to obtain thermodynamics is discussed in section 4. In section 5, using Mayer's convergence method with our new generating function, we study virial expansion and condensation. Finally, we conclude in section 6 that the new generating function may be directly lead to virial expansion and condensation at thermodynamic limit.  

\section{Mayer's theory of virial expansion:} 

Let us first survey Mayer's theory \cite{ma.1} based on Pathria's text book \cite{pa.1}.  The partition function for canonical ensemble (CE), $Q_N$, may be factorized into $Q_N = Z_N /(N! \lambda^{3N})$, where $Z_N$ is the configuration integral which involves the interaction effect or non-ideal behaviour and $\lambda$ is the thermal wave length. Using a diagrammatic analysis $Z_N$ may be expressed in terms of cluster integrals, 
\begin{equation}
Z_N = N! \lambda^{3N} \sum_{\{m_l\}}' \prod _{l=1}^{N} \frac{1}{m_l !} \left(\frac{b_l V}{\lambda^3} \right)^{m_l}  \,\, , \label{eq:zn} 
\end{equation}
where summation is with the restriction $\sum_{l=1}^{N} l m_l = N$. Various parameters are, $\lambda = $ thermal wavelength and $V = $ Volume and $b_l$ is the cluster integral defined as,
\begin{equation}
  b_l = \frac{1}{l! \lambda^{ 3(l-1)} V} \times \mbox{(sum of all possible l-clusters) } \,\,, \label{eq:b} 
\end{equation} 
such a way that $b_l$ is dimensionless and, in the limit $V-> \infty$ (thermodynamic limit), it is independent of size and shape \cite{pa.1}. This is true as long as $V/l$ is sufficiently large. But, for large clusters in which $l$ is about the value $N$, there will be a limit on specific volume $v=V/N$, called $v_f$ in Mayer's work, below which $b_l$ is not volume independent as discussed in in Ref. \cite{ma.1}. J. Mayer and M. Mayer related  $v_f$ to volume per molecule of condensed phase. Recently, a detailed volume dependence of $b_l$ and it's consequence was studied by Ushcats \cite{us.1}. Of course, there are different definitions of $b_l$ in the literature. In Mayer's work \cite{ma.1}, it was defined without $\lambda$ factor and recently, in Ushcats work \cite{us.1}, $\lambda^3$ in above Eq. (\ref{eq:b}) is replaced by volume $V$. However, this definition of $b_l$ makes $b_l$ $-> 0$ at thermodynamic limit. But the thermal wave length $\lambda$ is constant at given temperature and hence,  it may be  convenient to work with Pathria's dimensionless $b_l$.  The grand canonical partition function may be put into the form,
\begin{equation}
  Z =  \exp \left(\sum_{l\ge 1}  \frac{b_l V}{\lambda^3} z^l \right) \,\, ,
\end{equation}
and pressure,
\begin{equation}
  \frac{P}{k T} \equiv \frac{\ln Z}{V} = \frac{1}{\lambda^3} \sum_{l\ge 1}^{\infty} b_l z^l \,\, , \label{eq:ln}
\end{equation}
and average number density
\begin{equation}
  \frac{N}{V} \equiv \frac{1}{v} = \frac{1}{\lambda^3} \sum_{l\ge 1} l b_l z^l \label{eq:nu} \,\, .
\end{equation} 
Above two equation may be combined to get virial expansion, which is practically valid for dilute gas, as  
\begin{equation}
  \frac{P}{kT} = \frac{1}{v} \left( 1 - \sum_{k\ge 1} \frac{k}{k+1} \beta_k \left(\frac{\lambda^3}{v}\right)^k \right) \,\, ,\label{eq:vi} 
\end{equation}
where $\beta_k$ is known as irreducible cluster integral defined by 
\begin{equation}
    \beta_k = \frac{1}{k! \lambda^{3 k} V} \times \mbox{(sum of all possible irreducible (k+1)-clusters) } \,\,.
\end{equation}
Just like $b_l$, $\beta_k$ is also dimensionless and, in the limit $V-> \infty$, it is independent of size and shape \cite{pa.1}. The virial expansion Eq. (\ref{eq:vi}) may be derived as follows. First, by rewriting Eq. (\ref{eq:nu}) as 
\begin{equation}
  x(z) \equiv \frac{\lambda^3}{v(z)} = \sum_{l\ge 1} l b_l z^l  \label{eq:x} \,\, ,
\end{equation} 
and then, it may be inverted to get \cite{pa.1} 
\begin{equation}
  z = x \, e^{- \sum_{k\ge 1} \beta_k x^k} \,\, .
\end{equation}
Rewriting the equation for pressure, Eq. (\ref{eq:ln}), as 
\begin{equation}
  \frac{P}{k T} = \frac{1}{\lambda^3} \sum_{l\ge 1} b_l z^l = \int_0^z \frac{dz}{z} \frac{1}{v(z)} \,\, ,
\end{equation}
and on integration over $z$ gives virial expansion. $\beta_k$ may be related to $b_l$ using Lagrange's theorem \cite{pa.1} as,   
\begin{equation}
  b_l l^2 = \sum_{\{n_k\}}' \prod _{k=1}^{l-1} \frac{1}{n_k !} \left( l \beta_k  \right)^{n_k} \label{eq:bl} \,\,,
\end{equation}
where summation is with the restriction, 
\begin{equation}
 \sum_{k=1}^{(l-1)} k m_k = (l-1) \,\, .
\end{equation}
Comparing above Eq. (\ref{eq:bl}) with similar equation derived by Mayer and Mayer\cite{ma.1}, using their diagrammatic method, we may interpret $\beta_k$ as the irreducible cluster integral.  

\section{Another derivation of virial expansion:} 

In Mayer's theory, starting from $\sum_{l\ge 1} l b_l z^l = \frac{\lambda^3}{v}$ and inverting it, one gets virial expansion with $\beta_k$ related to $b_l$ through Eq. (\ref{eq:bl}). Following the recently developed formalism of Ushcats \cite{us.2} where one starts from Eq. (\ref{eq:bl}) and derives a relation, 
\begin{equation}
  \sum_{l\ge 1} l^2 b_l z^l = \frac{y}{1 - \sum_{k\ge 1} k \beta_k y^k} \,\, , \label{eq:l2b} 
\end{equation} 
where $z$ and $y$ are related by 
\begin{equation}
  z = y \, e^{- \sum_{k\ge 1} \beta_k y^k} \,\, .
\end{equation}
However, above equations differ from that of Ushcats by a factor $\frac{1}{V^k}$ inside the sum term. Instead of $\beta_k$ in the above equations and Eq. (\ref{eq:bl}), Ushcats has $\frac{\beta_k}{V^k}$ because of different definitions of cluster integral $b_l$. Eq. (\ref{eq:l2b}) is also derived in J. Mayer and M. Mayer \cite{ma.1}, but Ushacats procedure is simpler.  Eq. (\ref{eq:l2b}) may be further integrated with respect to $z$ to obtain,  
\begin{equation}
  \sum_{l\ge 1} l b_l z^l = y  \label{eq:y}  \,\, ,
\end{equation} 
and then, 
\begin{equation}
  \sum_{l\ge 1} b_l z^l = y \left( 1 - \sum_{k\ge 1} \frac{k}{k+1} \beta_k y^k \right) \label{eq:uv} \,\, .
\end{equation} 
Above equation has the same structure as that of virial expansion and hence Ushcats \cite{us.2} makes a comment about the similarity of the two equations. However, for GCE, by comparing  Eq. (\ref{eq:y}) and Eq. (\ref{eq:x}), and interpreting $z$ as fugacity, it naturally leads to the relation $y=\frac{\lambda^3}{v}$ and Eq. (\ref{eq:uv}) leads to  the virial expansion. Thus, we point out that the variable $y$ in our work, following Ushcats formalism, has the meaning of $\frac{\lambda^3}{v}$ in GCE.    

\section{Mayer's convergence method:} 

In the above two sections we saw that the virial expansion for GCE is derived by two different and opposite approaches, which is valid for $\frac{\lambda^3}{v} <1 $, but we get no information regarding phase-transition. To get few information about phase-transition or condensation of gas, one may use Mayers convergence method. We have from Eq. (\ref{eq:zn}), partition function, 
\begin{equation}
Q_N = \sum_{\{m_l\}}' \prod _{l=1}^{N} \frac{1}{m_l !} \left(\frac{b_l V}{\lambda^3} \right)^{m_l} \,\, ,\label{eq:qn} 
\end{equation}
which may be viewed as the expansion coefficients of a function $\exp (\sum_{l\ge 1} \frac{b_l V}{\lambda^3} z^l)$ in powers of $z$. That is, 
\begin{equation}
  F_M (z) = \exp \left( \sum_{l\ge 1} \frac{b_l V}{\lambda^3} z^l \right)  = \exp \left( N \sum_{l\ge 1} \frac{b_l v}{\lambda^3} z^l \right) = \sum_n a_n z^n \,\, . \label{eq:fm} 
\end{equation}
The coefficients $a_n$ for $n = N$ gives the partition function and hence $F_M (z)$ may be viewed as a generating function for $Q_N$. Extending to complex plane, above series may be viewed as Laurent series and the coefficient $a_N$ is 
\begin{equation}
  a_N = \frac{1}{2 \pi i} \int \frac{dz'}{z'^{N+1}} F_M(z') \,\, .
\end{equation}

Mayer's convergence method is based on Cauchy-Hadamard theorem \cite{ma.1} according to which the coefficient of a series expansion may be related to the radius of convergence of the series. Hence we consider a series, 
\begin{equation}
  H_0^0 (z,vb) \equiv \sum_{N=1}^{\infty} Q_N z^N = \sum_{N=1}^{\infty} a_N z^N 
\end{equation}
which on simplification, by sum and integration \cite{ma.1}, leads to 
\begin{equation}
H_0^0 (z,vb) = \frac{\sum_{l\ge 1} \frac{v}{\lambda^3}l b_l y^l}{1- \sum_{l\ge 1} \frac{v}{\lambda^3}l b_l y^l} \,\, , \label{eq:h0a} 
\end{equation}
where $y$ is related to $z$ through the relation, $z = y \, e^{ - \sum_{l\ge 1} \frac{v}{\lambda^3} b_l y^l}$. From the series expansion for $H_0^0(z,vb)$, using Cauchy-Hadamard theorem \cite{ma.1}, radius of convergence is 
\begin{equation}
  R = \lim_{N->\infty} Q_N^{-1/N}  \,\, ,
\end{equation} 
from which we can get the Helmholtz free energy, $A = - k T\, \ln Q_N = N k T\, \ln R$, where $R$ is the the value of $z$ at the singularity of $H_0^0(z,vb)$ which is one of the radius of convergence, say, $R_1$. This singularity corresponds to the value $y = Y$ in the equation,
\begin{equation}
1- \sum_{l\ge 1} \frac{v}{\lambda^3}l b_l Y^l = 0 \,\, , \label{eq:si} 
\end{equation}
and we get 
\begin{equation}
  R_1 = Y \, e^{ - \sum_{l\ge 1} \frac{v}{\lambda^3} b_l Y^l} \,\, . \label{eq:r1a} 
\end{equation} 
Thus we get Helmholtz free energy, 
\begin{equation}
  A = N k T \left[ \ln Y - \frac{v}{\lambda^3} \sum_{l\ge 1} b_l Y^l \right] \,\, , \label{eq:am} 
\end{equation}
and using $P = - \frac{\partial A}{\partial V}$, pressure is, 
\begin{equation}
  \frac{P}{k T} = \frac{\ln Z}{V} = \frac{1}{\lambda^3} \sum_{l\ge 1} b_l Y^l \,\, . \label{eq:pm}
\end{equation} 
Or, using the earlier result Eq. (\ref{eq:uv}),
\begin{equation}
   \frac{P}{k T} = \frac{1}{\lambda^3} y \left( 1 - \sum_{k \ge 1} \frac{k}{k+1} \beta_k y^k \right) \label{eq:mv} \,\, ,
\end{equation} 
where y is determined from the relation,
\begin{equation}
  Y = y \, e^{- \sum_{k\ge 1} \beta_k y^k} \,\, .
\end{equation}
At the same time Eq. (\ref{eq:y}) is
\begin{equation}
  \sum_{l\ge 1} l b_l Y^l = y  \label{eq:y2} 
\end{equation} 
which on comparison with the singularity relation, Eq. (\ref{eq:si}), gives, $y=\frac{\lambda^3}{v}$. Hence Eq. (\ref{eq:mv}) reduces to virial expansion, Eq. (\ref{eq:vi}). The chemical potential $\mu = \frac{\partial A}{\partial N}$ gives the fugacity $z=Y =  \frac{\lambda^3}{v} \, e^{- \sum_{k \ge 1} \beta_k (\frac{\lambda^3}{v})^k} $. 

Following the arguments of J. Mayer and M. Mayer \cite{ma.1},  $H_0^0 (z,vb)$ has also other singularities when $\sum_{l\ge 1} l b_l Y^l$ is singular or a general expression $\sum_{l\ge 1} l^n b_l Y^l$ is singular, where $n$ is an integer. For $n=2$ we have from Ushcats results, Eq. (\ref{eq:l2b}),
\begin{equation}
  \sum_{l\ge 1} l^2 b_l Y_i^l = \frac{y_i}{1 - \sum_{k\ge 1} k \beta_k y_i^k} \,\, , \label{eq:l2bn} 
\end{equation} 
 has singularities at $\sum_{k\ge 1} k \beta_k y_2^k = 1$  and $\sum_{k\ge 1} k \beta_k y_3^k$ itself is singular \cite{ma.1}. Here $Y_i$ and $y_i$ related as,
 \begin{equation}
  Y_i = y_i \, e^{- \sum_{k\ge 1} \beta_k y_i^k} \,\, , \label{eq:Yy} 
\end{equation}
where $i=2$ or $3$.
Let us denote corresponding radius of convergence be $R_2$ and $R_3$. That is, 
\begin{equation}
  R_i = Y_i \, e^{ - \sum_{l\ge 1} \frac{v}{\lambda^3} b_l Y_i^l} \,\, . \label{eq:ri} 
\end{equation} 
We have got the virial expansion by considering the first radius of convergence $R_1$ which is true for large $v$ or dilute gas. As $v$ decreases $R_1$ increases and it is possible that $R_2$ or $R_3$ may be lesser than $R_1$ and thus, $R_2$ or $R_3$ is the appropriate radius of convergence in Cauchy-Hadamard theorem. The singularity conditions to get $R_2$ or $R_3$, $\sum_{l\ge 1} l b_l Y^l =$ singular, and Eq. (\ref{eq:Yy}) are independent of specific volume $v$. Hence, $R_2$ or $R_3$ independent of $v$ and Helmholtz free energy, say, for $R_2$ is, 
\begin{equation}
  A = N K T \ln R_2 = N k T \left[ \ln Y_2 - \frac{v}{\lambda^3} \sum_{l\ge 1} b_l Y_2^l \right] \,\, , \label{eq:ami} 
\end{equation}
which depends on $v$, but pressure, using $P = - \frac{\partial A}{\partial V}$, we get, 
\begin{equation}
  \frac{P}{k T} = \frac{1}{\lambda^3} \sum_{l\ge 1} b_l Y_2^l = \frac{1}{\lambda^3} y_2 \left( 1 - \sum_{k \ge 1} \frac{k}{k+1} \beta_k y_2^k \right) \,\, , \label{eq:pmi}
\end{equation} 
where $y_2$ is the solution of the singularity condition $\sum_{k\ge 1} k \beta_k y_2^k = 1$. Thus, pressure is independent of specific volume and it may corresponds to condensation as suggested by J. Mayer and M. Mayer \cite{ma.1} and recently, by Ushcats \cite{us.1}. If $R_3$ is less than $R_1$ and $R_2$, for which $\sum_{k\ge 1} k \beta_k y_3^k$ is singular, it is not clear how to solve to get $y_3$.     

\section{Mayer's convergence method using a new generating function:} 

In the earlier section we used the fact that the expression for $Q_N$ in terms of $b_l$, Eq. (\ref{eq:qn}), and expression for $b_l l^2$  in terms of $\beta_k$, Eq. (\ref{eq:bl}), has a similar structure. Hence, one obtains $\sum_l b_l z^l$, $\sum_l l b_l z^l$,  etc. related to $\beta_k$ and discussed virial expansion and condensation. 

But, here, using Ushcats procedure \cite{us.1, us.2},  we may express $\sum_l b_l z^l$ in terms of $\beta_k$ from the start, Eq. (\ref{eq:fm}), and obtain a new generating function for $Q_N$ in terms of $\beta_k$ as 
\begin{equation}
  F_B(y) = (1 -\sum_{k\ge 1} k \beta_k y^k) \,\,  e^{ N \left[ \frac{v}{\lambda^3} y \left( 1 - \sum_{k\ge 1} \frac{k}{k+1} \beta_k y^k \right) + \sum_{k\ge 1} \beta_k y^k \right] } = \sum_n a_n y^n  \,\, ,
\end{equation}
instead of $F_M (z) = e^{ N \sum_{l\ge 1} \frac{b_l v}{\lambda^3} z^l} = \sum_n a_n z^n$ and $a_n = Q_N$ for $n = N$. It differs from that of Ushcats \cite{us.1} in number of ways. First, as previously indicated, there is no ($\frac{1}{V^k}$) factor inside the sum. Secondly, there is a factor $\frac{N v}{\lambda^3}$ outside the sum because of the different definition of $b_l$ which resulted in different generating function $F_M(z)$ compared to that of Ushcats. Thirdly, our new generating function is in a suitable form for applying Mayers convergence method to discuss condensation. However, note that we may obtain Ushcats generating function from our generating by replacing the variable $y$ by $\frac{\lambda^3}{V} y$.  Using our generating function and following the convergence method of Mayers as discussed in the earlier section, we get a very simple relation after some algebra,

\begin{equation}
  H_0^0 (y,v) \equiv \sum_{N=1}^{\infty} Q_N y^N = \sum_{N=1}^{\infty} a_N y^N 
  = \frac{\frac{v}{\lambda^3} Y}{1 -\frac{v}{\lambda^3} Y} \,\, , \label{eq:h0b}
\end{equation}
where 
\begin{equation}
  a_N = \frac{1}{2 \pi i} \int \frac{dy'}{y'^{N+1}} F_B(y') \,\,,
\end{equation}
and $Y$ is related to $y$ through the equation 
\begin{equation}
  y = Y \, e^{ - \left[ \frac{v}{\lambda^3} Y \left( 1 - \sum_{k\ge 1} \frac{k}{k+1} \beta_k Y^k \right) + \sum_{k\ge 1} \beta_k Y^k \right] } \,\, .
\end{equation}
Hence, the radius of convergence ($R_1$) is at $Y = \frac{\lambda^3}{v} \equiv Y_1$ and
\begin{equation}
  R_1 = y_{convergence} =  \frac{\lambda^3}{v} e^{ - \left( 1 + \sum_{k\ge 1} \frac{1}{k+1} \beta_k (\frac{\lambda^3}{v})^k \right) }\,\, .\label{eq:r1b} 
\end{equation} 
Note that Eqs. (\ref{eq:h0b}) and (\ref{eq:r1b}) are analogous to Eqs. (\ref{eq:h0a}) and (\ref{eq:r1a}) in Mayer's method, but here they are directly related to irreducible cluster integrals. From $R_1$ we may very easily get $A$ and then $P$ in the form of virial expansion by following the steps of Mayer's theory, analogous to  Eqs. (\ref{eq:am}) and (\ref{eq:pm}). Besides this singularity, $H_0^0$ has another singularity for $\sum_{k\ge 1} k \beta_k Y_2^k = 1$ and corresponding radius of convergence $R_2$. As pointed out by Mayers, there may be other singularities when $\sum_{k\ge 1} k \beta_k Y_3^k$ is singular and corresponding radius of convergence, say, $R_3$ . This is because $H_0^0$ is not an analytic function and it's higher order derivative has singularities other than first one, $\frac{v}{\lambda^3} Y_1 = 1$. But, $Y_2$ and $Y_3$,  corresponding to other two singularities, are independent of $v$. For low density such that $R_1$ is less than other two, we have pure gaseous state and virial expansion for pressure. $R_1 < R_2$ corresponds to $\sum_{k\ge 1} k \beta_k Y_1^k = \sum_{k\ge 1} k \beta_k (\frac{\lambda^3}{v})^k < 1$. But as density increases, after some critical value, say, $1/v_s$, $R_2$ or $R_3$ may be lower than $R_1$ and condensation starts. For $R_1 = R_2$ we have the condition $\sum_{k\ge 1} k \beta_k (\frac{\lambda^3}{v})^k = 1$ which gives a critical specific volume $v_s$ above which virial expansion is valid. For $v<v_s$ thermodynamics are derivable from $R_2$ which is independent of $v$ and hence as discussed in the earlier section, pressure is independent of $v$ and is determined by  on the critical volume $v_s$.This $v_s$ happens to be the solution of $(\frac{\partial P}{\partial v})_T =0$ in virial expansion, Eq.(\ref{eq:vi}), which is called specific volume of saturated vapour pressure \cite{ma.1}. Of course, as pointed out by Mayers and recently by Ushcats, there is a temperature range for which  $(\frac{\partial P}{\partial v})_T =0$, but properties of such a system are not associated with condensation.  It all consistent with Mayers observation, but here it follows from our new generating function. It also clarifies why the condition $\sum_{k\ge 1} k \beta_k (\frac{\lambda^3}{v})^k < 1$ is needed for the validity of virial expansion.  

Hence, all the properties such as virial expansion, condensation, etc., discussed by Mayers \cite{ma.1}, may be re-derived using a new generating function. The advantage here is that the generating function is directly expressed in terms of $\beta_k$ and hence, $Q_N$ depends on virial coefficients and as pointed out by Ushcats \cite{us.1}, one may study even a system with finite $N$ by computational methods. But in Mayers formulation, generating function depends on $b_l$ which need to be expressed in terms of $\beta_k$ to get virial expansion.  

Thus we see that Mayer's convergence method with our new generating function may be used to evaluate $Q_N$ instead of solving complicated Eq.(\ref{eq:qn}). Similarly, we may apply the convergence method to the evaluation of $b_l$, instead of Eq.(\ref{eq:bl}), which has the similar structure as $Q_N$ and obtain the criteria for the existance of clusters with large $l$ which may corresponds to condensation. Hence let us apply the same procedure of Mayer's convergence method. $l^2 b_l $ may be viewed as the expansion coefficients of a function, 
\begin{equation}
  F_U (z) = \exp( \sum_{k \ge 1} l \beta_k z^k) = \sum_n a_n z^n \,\, . \label{eq:fu} 
\end{equation}
The expression for $l^2 b_l$, Eq. (\ref{eq:bl}), is the coefficient $a_n$ for $n=l-1$. 
Ushcats \cite{us.2} had already evaluated the series, Eq. (\ref{eq:l2b}),        
\begin{equation}
  \sum_{l\ge 1} l^2 b_l z^l = \frac{y}{1 - \sum_{k\ge 1} k \beta_k y^k} \,\, , 
\end{equation} 
and the radius of convergence in above series is related to $l^2 b_l$ as  
\begin{equation}
  R_b = \lim_{l->\infty} (l^2 b_l)^{-1/l}  \,\, . 
\end{equation} 
$R_b$ may be obtained from the divergence of the right hand side of Eq. (\ref{eq:l2b}). That is, $\sum_{k\ge 1} k \beta_k Y_b^k = 1$, which is the same condition as that for the second divergence of $H_0^0$ in the case of partition function. Therefore, $Y_b$ is the same as $Y_2$ which is a condition for condensation or larger cluster formation. Thus we obtained the same condition for condensation by three different approaches. The Mayer's convergence method for the generating function of $Q_N$ leads to $\sum_{k\ge 1} k \beta_k Y_2^k = 1$, convergence method to $b_l l^2$ generating function leads to $\sum_{k\ge 1} k \beta_k Y_b^k = 1$ and $(\frac{\partial P}{\partial v})_T =0$ of virial expansion leads to  $\sum_{k\ge 1} k \beta_k (\frac{\lambda^3}{v})^k = 1$. Note that there is one more approximate method to evaluate $Q_N$ from  Eq.(\ref{eq:qn}), and $b_l$ from Eq.(\ref{eq:bl}), based on variation method given in Ref. \cite{ma.1} and similar conditions for condensation and virial expansion, as discussed here, are obtained. 
  
\section{Conclusions:} 

We derived a new generating function for the canonical partition function. Using this generating function, we reanalyzed the Mayers theory of classical cluster expansion and studied Mayers theory of virial expansion and condensation at thermodynamic limit. The new generating function depends on the irreducible cluster integral $\beta_k$ and hence directly leads to virial expansion, condensation, etc. Various issues like low density limit, problem of divergences, etc. of virial expansion and condensation were clarified and consistent with Mayer's theory. As pointed out by Ushcats, equation of state in terms of irreducible cluster integral for finite $N$ system may be easily computed. In Mayers calculations the generating function may be simpler, but depends on cluster integral $b_l$ and hence the dependence of equation of state on irreducible cluster integral or virial expansion limits are not direct. One needs to use another relation connecting the cluster integrals $b_l$ and $\beta_k$.

\end{document}